\documentclass[reprint,aps,prl,superscriptaddress,altaffiliation,superscriptaddress]{revtex4-2}
\usepackage{stmaryrd}
\usepackage{amsmath}
\usepackage{amssymb}
\usepackage{graphicx}
\usepackage{dcolumn}
\usepackage{bm}
\usepackage{tabularx}
\usepackage{color}

\begin{document}
	\begin{titlepage}
		\title{Uncovering the Fourier Structure of Wavefunctions in Semiconductors}
		\author{Yunfan Liang}
		\affiliation{Department of Physics, Applied Physics and Astronomy, Rensselaer Polytechnic Institute, Troy, NY, 12180, USA}
		\author{Damien West}
		\affiliation{Department of Physics, Applied Physics and Astronomy, Rensselaer Polytechnic Institute, Troy, NY, 12180, USA}
		\author{Shengbai Zhang}
		\affiliation{Department of Physics, Applied Physics and Astronomy, Rensselaer Polytechnic Institute, Troy, NY, 12180, USA}
		\email{zhangs9@rpi.edu}
		\date{\today}
		
		\begin{abstract}
		Symmetry is at the heart of material properties. Symmetry of the Bravais lattice defines the degeneracy of planewaves, upon which atomic symmetry determines interaction potentials which may lift such degeneracies. This results in wavefunctions which are single planewaves throughout the Brillouin zone (BZ), except in the vicinity of lifted degeneracies. This great simplification allows for determination of optical properties from a handful of planewaves and a single transition. Further, it reveals that nonlinear optical response arises from higher order degeneracy along lines/points in the BZ.
		\end{abstract}
		
		\maketitle
		\draft
		\vspace{2mm}
	\end{titlepage}

	
	Optical properties are fundamental for many materials applications \cite{cohen2012electronic,wang2012electronics,guo2017recent}, yet they are complex and involve  transition dipole matrix elements between different states \(\langle\psi_{n,\mathbf{k}}\left| \mathbf{r} \right|\psi_{n^{'},\mathbf{k}}\rangle\) which span the energy spectrum and are over the entire Brillouin zone (BZ). Although first-principles methods can directly calculate these matrix elements
	\cite{nunes2001berry,souza2002first,gajdovs2006linear,ambrosch2006linear}, the complexity makes it challenging to reveal the underlying physics which control these properties. Much of the current understanding is still based on the joint density of states \cite{amiotti1990optical,o1995optical,o1997relationship}, wherein the matrix elements are set to unity.
	
	Beyond dipole matrix elements, any perturbative treatment of the system will involve transition matrix elements between states, which broadly affect the material characteristics, e.g., dielectric screening \cite{bertoni1974dielectric,hybertsen1987ab}, radiative \cite{dumke1957spontaneous,dmitriev1999rate} and non-radiative recombination \cite{laks1990accurate,kioupakis2015first}, free-carrier absorption \cite{kioupakis2010free}, electron-phonon coupling \cite{giustino2017electron}, and therefore carrier scattering \cite{murphy2008first}. In the many-body GW quasiparticle calculations, the difficult to converge Coulomb hole involves matrix elements of the form \(\langle\psi_{n,\mathbf{k}}\left| e^{i\left( \mathbf{q} + \mathbf{G} \right) \cdot \mathbf{r}} \right|\psi_{n^{'},\mathbf{k} - \mathbf{q}}\rangle\) where \(n'\) runs over unoccupied states \cite{zhang1989evaluation}.
	
	To develop physical insight into these matrix elements, and hence poorly understood physical properties such as the nonlinear susceptibility, it is critically important to gain an understanding of the wavefunction structure in Fourier space. In this regard, the simplest framework is to view the solid as a free electron gas (FEG) in the presence of a periodic atomic potential. Here, the Bravais lattice defines the FEG states and crystal symmetry imposes restrictions on non-vanishing Fourier components of the potential \(V(\mathbf{G})\). This so-called empirical pseudopotential method (EPM) has been shown to provide accurate descriptions of the band structures, deformation potentials and optical properties of bulk materials, alloys and nanostructures \cite{friedel1989local,zhang1993electronic,fischetti1996band,kim2024transferable}.
	
	In this paper, using Bravais lattice symmetry, we gain physical insight into the wavefunction, enabling simple and accurate analytical understanding of optical response in non-transition-metal solids. A key realization is that all optical transitions of the FEG are \emph{dark}. Hence, only transitions between states which contain common Fourier components can be optically active. Such mixing is most prominent when the potential of the solid breaks the degeneracy of the FEG, defined by the Bravais lattice. Taking Si as a prototype, we show that only the states derived from doubly degenerate FEG states within symmetry planes in BZ dominate the linear optical response. By considering only two planewaves (2G-model), we construct an analytic expression of the transition matrix elements containing only a single coupling parameter determined from first principles. \(\chi^{(1)}\) predicted from this model agrees well with first-principles results (see 'Method' section in supplementary information (SI) \cite{gonze2002first,gonze2009abinit,gonze2020abinit,perdew1996generalized,madelung2004semiconductors}) and the deviations are well understood by considering higher order degeneracy along symmetry lines. Further, it is precisely these lines of higher order degeneracy which give rise to nonlinear optical response. This is demonstrated for GaAs, where contributions from the vicinity of the $\Gamma$-$L$ lines dominate $\chi^{(2)}(\hbar \omega)$. This is a substantial advance in the current understanding of nonlinear optics which largely neglects the transition matrix element, focusing on the joint density of states.
	
	As a starting point for understanding the wavefunction structure in the BZ, we investigated the 1D Kronig-Penney (KP) model \cite{singh1983kronig} for a periodic potential \(V(x) = \alpha\delta(x - nL)\), where \(\alpha\) is the potential strength and \(L\) is the periodicity. The KP model can be solved analytically in Fourier space as detailed in 'KP model' section in SI. The band structures of the FEG and KP models are shown in Fig. 1(a), in gray and color, respectively. The FEG eigenstates are planewaves and the corresponding band structure is a folded parabola, leading to double degeneracies at the symmetry points, i.e., \(\Gamma\) and \(\pm X\). After applying the potential \(V(x)\), these degenerate states are split, opening up band gaps. Away from these k-points, however, the bands quickly reduce to the FEG band structure of single planewaves. This can directly be seen by examining the projection of each band onto the i\textsuperscript{th} planewave, shown by the color in Fig. 1(a).
	\begin{figure}[tbp]
		\includegraphics[width=\columnwidth]{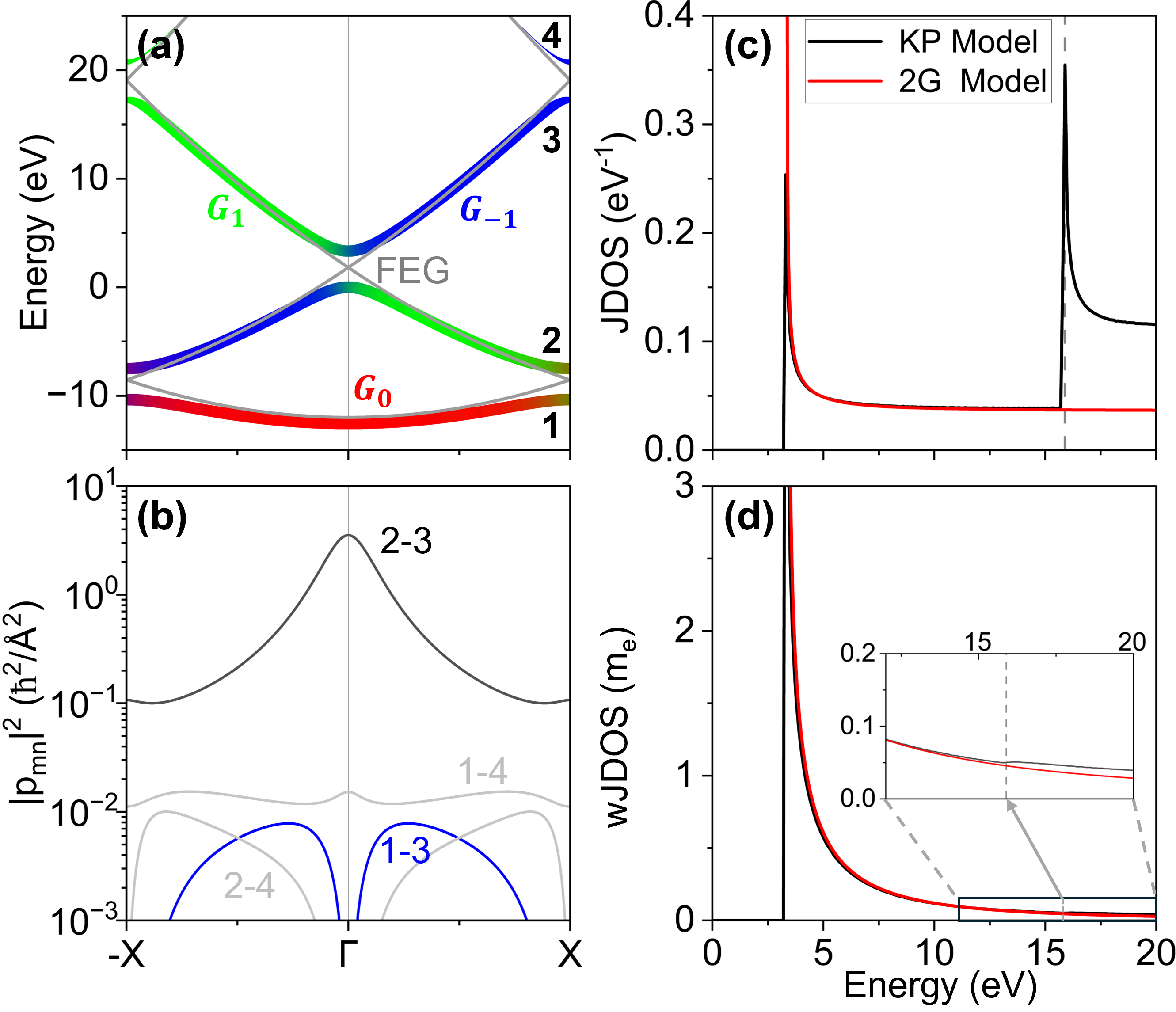}
		\caption{\label{fig:fig1} (a) Shows the energy dispersion, in which we assume the first two bands are fully occupied, while all others are fully empty in line with later discussion for Si. The average potentials for FEG and KP model are aligned. The VBM is set to 0 eV. (b) Shows the momentum matrix element squared. (c) Shows the joint density of states (JDOS), and (d) shows $ \left| p_{mn,k} \right|^{2}$ weighted JDOS (wJDOS).}
	\end{figure}
	
	To investigate the optical properties, we examine the dipole transition matrix $(r_{mn,k})$, which for solids is determined from the momentum matrix element  \(p_{mn,k}\) \cite{sipe1993nonlinear,sharma2004second,ambrosch2006linear},
	 with $r_{mn,k} = \frac{i \hslash}{m_{e}}\frac{p_{mn,k}}{E_{n,k} - E_{m,k}}$. As planewaves are eigenfunctions of the momentum operator, for the FEG, $p_{mn,k}$ vanishes for all $m \neq n$, indicating all transitions are forbidden. Hence, in solids, \(p_{mn,k}\) can only be nonzero if states m and n share at least one common planewave, \(\mathbf{G}_{i}\). Therefore, for the KP model, the strength of dipole transition measured by \(\left| p_{mn,k} \right|^{2}\) is negligibly small, except near symmetry points of the BZ. For example, only the transitions between the 2\textsuperscript{nd} and 3\textsuperscript{rd} band near $\Gamma$ in Fig. 1(b) have substantial \(\left| p_{mn,k} \right|^{2}\). This suggests that the optical properties can be understood simply in terms of the two planewaves, \textbf{G}\textsubscript{1} and \textbf{G}\textsubscript{-1\textbf{, }}associated with this transition. Further, it highlights the importance of the transition matrix elements whose information is lacking in the joint density of states (JDOS), Fig. 1(c), which simply measures the density of transitions in an energy window.
	
	To incorporate the strength of these transitions, here we introduce the transition matrix element weighted joint density of states (\(wJDOS)\) shown in Fig. 1(d),
	\begin{equation} \label{wJDOS_Def}
		\begin{gathered}
			wJDOS(E) = \sum_{k}^{}{\sum_{m,n}^{}{\delta\left( E_{m,k} - E_{n,k} - E \right)\left| p_{mn,k} \right|^{2}}} \\
			= \sum_{k}^{}{f(E,\mathbf{k})}
		\end{gathered}
	\end{equation}
	where the integral of \(wJDOS(E)E^{- 1}(E^2 - \hslash^2\omega^2)^{- 1}\) is directly proportional to the susceptibility,
	\(\chi^{(1)}(\hslash\omega)\). For KP, the most striking difference between the JDOS and wJDOS seen in Figs. 1(c) and (d) is that the prominent second peak near 15eV in the JDOS has been drastically reduced in the wJDOS, seen in the inset of Fig. 1(d). This indeed indicates that \(\chi^{(1)}\) can be understood simply from the primary peak associated with the transition between the 2\textsuperscript{nd} and 3\textsuperscript{rd} bands. To understand this wJDOS, we construct a simple two band Hamiltonian (describing the 2\textsuperscript{nd} and 3\textsuperscript{rd} bands) with the two planewaves \(G_{1}\) and \(G_{- 1}\), where the coupling is half the bandgap.
	\begin{equation} \label{2G_Ham}
			H = \begin{pmatrix}
				\frac{\hslash^{2}}{2m_{e}}\left( k + G_{1} \right)^{2} & \frac{E_{g}}{2} \\
				\frac{E_{g}}{2} & {\frac{\hslash^{2}}{2m_{e}}\left( k + G_{- 1} \right)}^{2} \\
			\end{pmatrix}.
		\end{equation}
		The results of this 2G-model are shown in red curves in Figs. 1(c) and (d) (see "2G-model" section in SI). While the second peak in the JDOS is not represented in this model, we can see the wJDOS (and hence \(\chi\)) is very well represented over the entire energy range.
		
		From the EPM perspective, the electronic states of real semiconductors can be understood as a set of planewaves that are coupled by a periodic potential which is expressed as a Fourier series
		\begin{equation}\label{Poten}
			V\left( \bm{r} \right) = \sum_{{\bm{G}}_{s}}^{}{V_{{\bm{G}}_{s}}e^{i{\bm{G}}_{s} \cdot \bm{r}}},
		\end{equation}
		where, \(V_{{\bm{G}}_{s}}\) describes the direct coupling between two planewaves whose G-vectors differ by
		\({\bm{G}}_{s}\), which decays rapidly with the increase of \(\left| {\bm{G}}_{s} \right|\). For example, only 3 leading terms \(V_{\left| {\bm{G}}_{s} \right| = \sqrt{3},\ \sqrt{8},\ \sqrt{11}}\) are needed to accurately describe the Si band structure \cite{zhang1993electronic}. Since the FEG states are also a good reference to understanding real semiconductors, the insight of wavefunction structure in Fourier space obtained from 1D model can be naturally extended to real semiconductors.
		\begin{figure}[tbp]
			\includegraphics[width=\columnwidth]{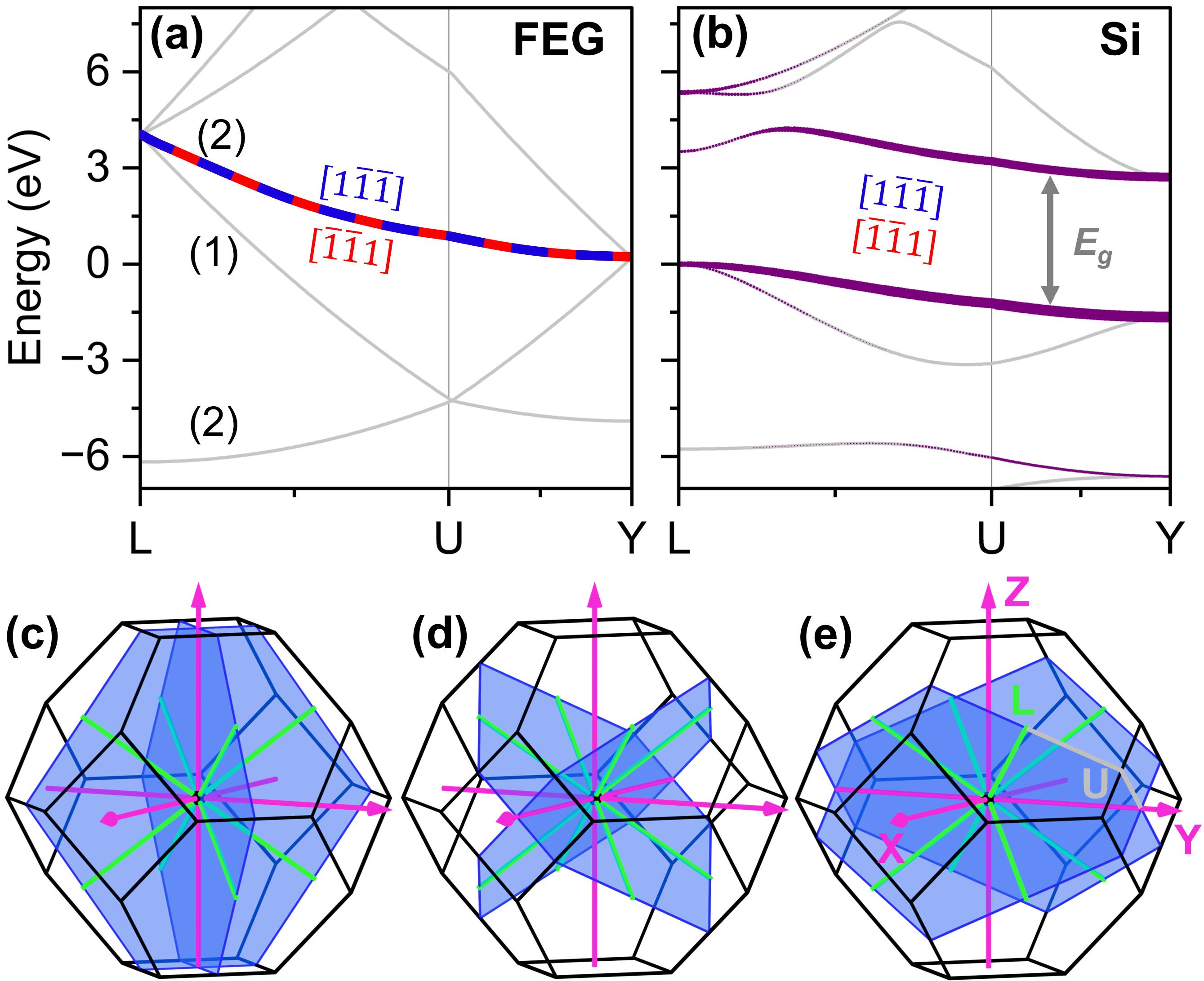}
			\caption{\label{fig:fig2} The band structure of (a) FEG and (b) Si with the wavefunction projection on Fourier component \(\lbrack1\overline{1}\overline{1}\rbrack\) (blue) and \(\lbrack\overline{1}\overline{1}1\rbrack\) (red). In panel (b) the purple color indicates the uniform mixing between these two components. (c-e) sketches the \(\sqrt{8}\)-planes, which are (c) \(k_{x} = \pm k_{y}\) (d) \(k_{y} = \pm k_{z}\) and (e) \(k_{z} = \pm k_{x}\)}
		\end{figure}
		
		In Fig. 2, we compare the FEG band structure to that of Si calculated using DFT. Surprisingly, these band structures look quite similar, and the details of the Si band structure can be primarily understood from the splitting of degeneracies in the FEG. A key distinction from the 1D case is that in 3D the degeneracy of the FEG is not limited to high symmetry points, but states can also be degenerate along high-symmetry lines or within planes of the BZ. Indeed, along the \(L \rightarrow U \rightarrow Y\) path shown in Fig. 2, the \({\bm{G}}_{n1} = \lbrack1\overline{1}\overline{1}\rbrack\)
		and \({\bm{G}}_{n2} = \left\lbrack \overline{1}\overline{1}1 \right\rbrack\) planewaves are degenerate for the FEG. However, for Si, in addition to
		the splitting seen at high symmetry points, e.g. L and U, this doubly degenerate band splits along the entire path, leading to the opening of the gap between occupied and unoccupied states which is responsible for the semiconducting nature of Si.
		
		The splitting of this doubly degenerate band can be understood from the EPM description of Si, in which the \(\sqrt{8}\) term of the potential directly couples \(\bm{G}_{n1}\) and \(\bm{G}_{n2}\), as \(\left| \bm{G}_{n1} - \bm{G}_{n2} \right| = \sqrt{8}\). As the gap originates from this splitting, the highest valence band and lowest conduction band have Fourier component projections, Fig. 2(b), containing equal weights of \(\bm{G}_{n1}\) and \(\bm{G}_{n2}\), similar to the mixing observed at \(\Gamma\) in Fig. 1(a). This is especially important in an optical context, as it is the states, which arise from such splitting, have substantial strength of dipole transition.
		
		To understand the optical properties, we focus on states derived from doubly degenerate (n=2) FEG states, dominated by two Fourier components (i.e., \({\bm{G}}_{n1}\) and \({\bm{G}}_{n2}\)), as these states are degenerate along an entire plane in the BZ, and hence have the largest weight. To determine these we consider, 
		(1) The Bravais lattice, which determines the set of \({\bm{G}}\) vectors, and hence the planar degeneracy of the FEG, where
		\begin{subequations}
			\begin{equation}
				\left( \bm{k} + {\bm{G}}_{n1} \right)^{2} = \left( \bm{k} + {\bm{G}}_{n2} \right)^{2} \\ \label{Deg_FEG}
			\end{equation}
		(2) Next, the atomic symmetry determines the non-vanishing components of the potential, and hence dictates which of these states split. For Si, this means:
			\begin{equation}
				\left| {\bm{G}}_{n1} - {\bm{G}}_{n2} \right| = \sqrt{3},\sqrt{8},or\ \sqrt{11} \label{Si_EPM}
			\end{equation}
		\end{subequations}
		(3) For linear optical response, only the dipole transitions from
		occupied to unoccupied states affect the properties, as such the split degeneracy must straddle the band gap. The doubly degenerate FEG states that satisfies all these requirements are listed in Tab. S1 in SI, which shares the common features of
		\(\left| {\bm{G}}_{n1} - {\bm{G}}_{n2} \right| = \sqrt{8}\) and
		\(\left| {\bm{G}}_{n1} \right| = \left| {\bm{G}}_{n2} \right|\). For these FEG states, the set of k points satisfying Eq. \eqref{Deg_FEG} forms 6 planes, denoted the \(\sqrt{8}\)-planes, in the first BZ, which are \(k_{x} = \pm k_{y}\) in Fig. 2 (c), \(k_{y} = \pm k_{z}\) in Fig. 2(d) and \(k_{z} = \pm k_{x}\) in Fig. 2(e).
\begin{figure}[tbp]
	\includegraphics[width=\columnwidth]{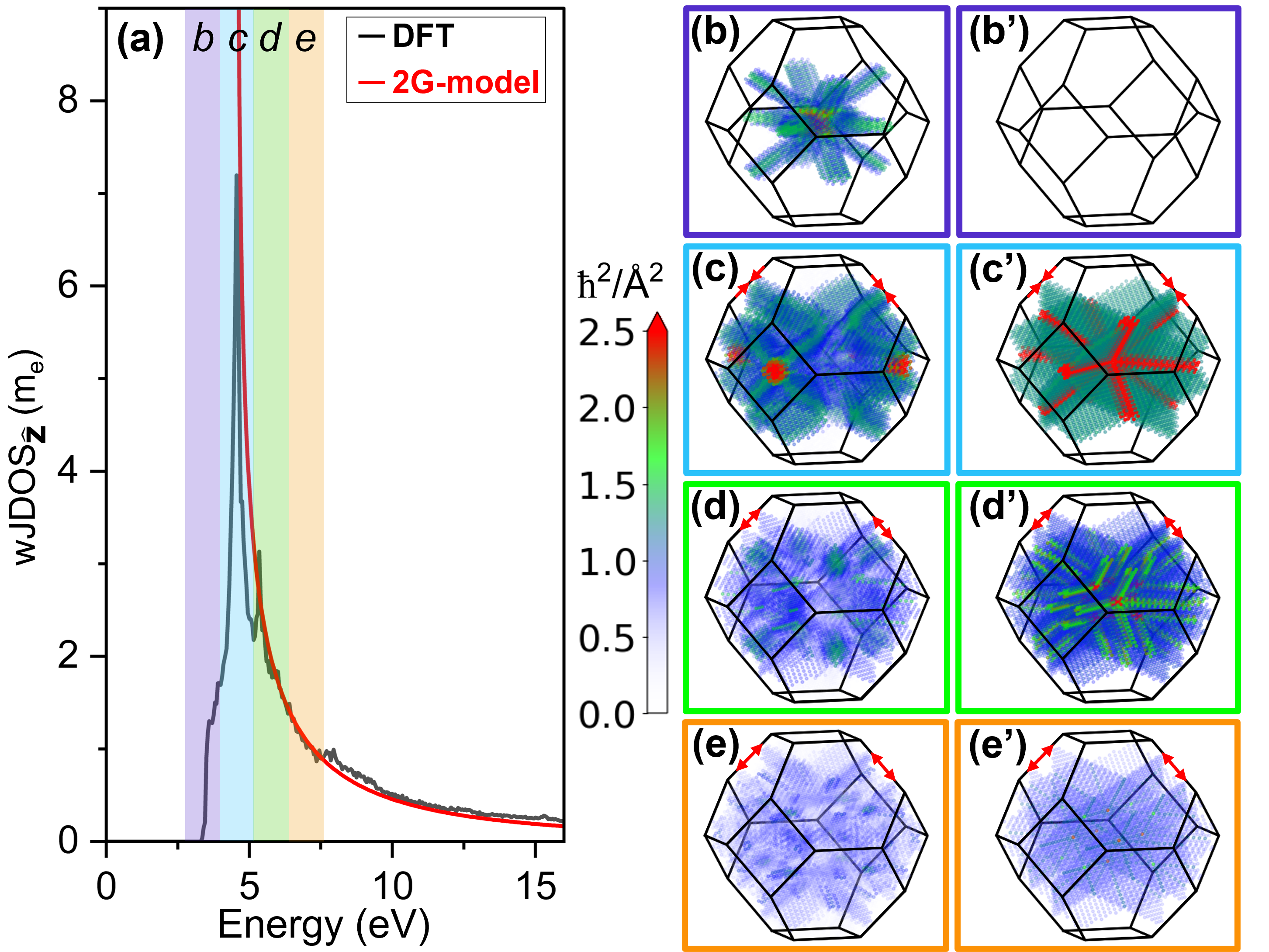}
	\caption{\label{fig:fig3} (a) shows the $wJDOS_{\hat{z}}$ of bulk Si from DFT calculation (black) and 2G-model (red). (b-e) and (b'-e') show the integral of $f(E, \bm{k})$ in Eq. \eqref{wJDOS_Def} over $E$ within each energy window indicated in (a) for each k point from (b-e) DFT calculation and 2G-model (b'-e').}
\end{figure}		
		After coupling, the FEG states (planewaves) mix to form Si states. Within these \(\sqrt{8}\)-planes, the splitting of the double degeneracy leads to states where \({\bm{G}}_{n1}\) and
		\({\bm{G}}_{n2}\) have equal weights. Using the same simple model presented in Eq. \eqref{2G_Ham} (see "2G-model in 3D" section of the SI for details),
		\begin{subequations}
			\label{E_P_3D}
			\begin{equation}
				\left| \left( {\bm{p}}_{mn,\bm{k}} \right)_{z} \right|^{2} = \hslash^{2}\frac{A^{2}{\left( {\bm{G}}_{n1} - {\bm{G}}_{n2} \right)_{z}}^{2}}{E^{2}} \label{P_3D}
			\end{equation}
			\begin{equation}
				E(\bm{k}) = \sqrt{\left( \frac{\hslash^{2}}{m_{e}} \right)^{2}\left( \bm{k} \cdot \hat{\bm{n}} \right)^{2}\ \left| {\bm{G}}_{n1} - {\bm{G}}_{n2} \right|^{2} + 4A^{2}}, \label{E_3D}
			\end{equation}
		\end{subequations}
		where E is the energy difference after splitting, A is the coupling constant and \(\hat{\bm{n}}\) is the normal vector of the \(\sqrt{8}\)-plane. When moving away from the plane, the energy difference increases while the strength of dipole transition decreases. Focusing on light polarized in the z-direction, from Eq. \eqref{E_P_3D},
		\begin{equation} \label{wJDOS_3D}
			{wJDOS}_{\hat{z}} = 8m_{e}\frac{A^{2}}{E^{2}}\sqrt{\frac{E^{2}}{E^{2} - 4A^{2}}},
		\end{equation}
		where the \(\hat{z}\) indicates that the JDOS is weighted by
		\(\left| \left( {\bm{p}}_{mn,\bm{k}} \right)_{z} \right|^{2}\),
		and the coupling constant A is determined from the energy splitting of the states in DFT ($A = E_g / 2$), as shown in Fig. 2(b). Similar to the KP model in 1D, here we see that the 2G-model faithfully represents the major peak and tail of the 3D ${wJDOS}_{\hat{z}}$ of Si determined by DFT in Fig. 3(a).

		To gain further insight, \(f(E,\mathbf{k})\), defined in Eq. \eqref{wJDOS_Def}, is integrated over the energy windows labeled b, c, d and e in Fig. 3(a) and shown in Figs. 3(b-e) for DFT and in Figs. 3(b'-e') for the 2G-model. While (b) and (b') shows differences, which will be discussed later, both Figs. 3(c) and (c') have the major contribution from the four \(\sqrt{8}\)-planes associated with \({(\mathbf{p}_{mn,k})}_{z}.\). Here it can be seen that it is indeed these planes which give rise to the major peak in the ${wJDOS}_{\hat{z}}$, seen in energy window c. As the energy window is shifted to higher energies, this corresponds to a larger energy separation between states in the 2G-model (Eq. \eqref{E_3D}) which is associated with increasing the distance (\(\mathbf{k} \cdot \hat{n})\) from the \(\sqrt{8}\)-planes. As a result, Figs. 3(d) and (d') reveal that the major contributions originate from pairs of planes parallel to but offset from the \(\sqrt{8}\)-planes. As the energy further increases, the spacing between these planes grows and the magnitude of the matrix elements diminishes as shown in Figs 3(e) and (e').
		
	\begin{figure}[tbp]
	\includegraphics[width=\columnwidth]{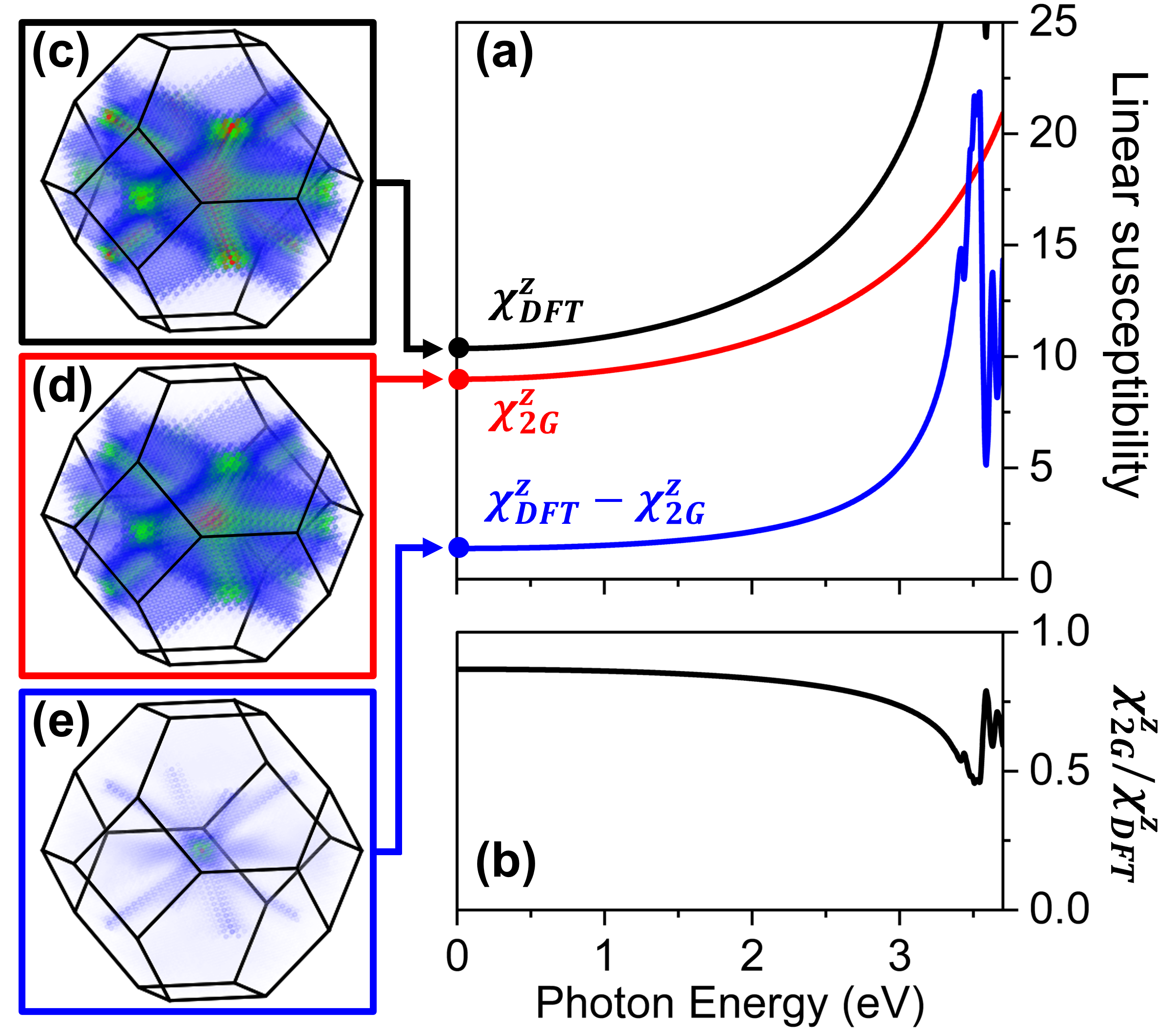}
	\caption{\label{fig:fig5} (a) shows the real part of frequency dependent z component of total susceptibility \(\chi_{DFT}^{z}(\omega)\) determined from DFT (black) the \(\chi_{2G}^{z}(\omega)\) determined from the 2G-model (red) and their difference (blue). (b) shows the ratio between \(\chi_{2G}^{z}(\omega)\) 	and \(\chi_{DFT}^{z}(\omega)\). (c-d) show that k-resolved susceptibility of (c) \(\chi_{DFT}^{z}(0,\mathbf{k})\), (d) \(\chi_{2G}^{z}(0,\mathbf{k})\) and (e) their difference.}
	\end{figure}

		As the 2G-model captures the essential physics of the splitting of the two-fold planar degeneracy, the major peak and tail of the  $wJDOS_{\hat{z}}$ are well represented. Further, in the \(\hslash\omega \rightarrow 0\) limit, it captures nearly 90\% of the contribution to linear optical susceptibility as depicted in Fig. 4. While the 2G-model focuses on degenerate planes, we note that at the intersection of these planes, there are lines of higher degeneracy involving more than two planewaves. The effect of neglecting these higher order degeneracies can be seen in Fig. 3(a), where a small shoulder in energy window b, causing difference between Figs. 3(b) and (b'), and a small secondary peak in energy window d. As detailed in Fig. S2 in SI, spatial decomposition of the $wJDOS_{\hat{z}}$ clearly shows that the region of the BZ near the $\Gamma$ to $L$ lines is responsible for both the shoulder and secondary peak while the $wJDOS_{\hat{z}}$ of the other region is accurately captured by the 2G-model. To describe the higher order degeneracy along symmetry lines requires an nG-model with n$>$2. A 6G-model effectively captures the key features of these higher order contributions to Si, III-V, and II-VI semiconductors, as detailed in the SI.
\begin{figure}[tbp]
	\includegraphics[width=\columnwidth]{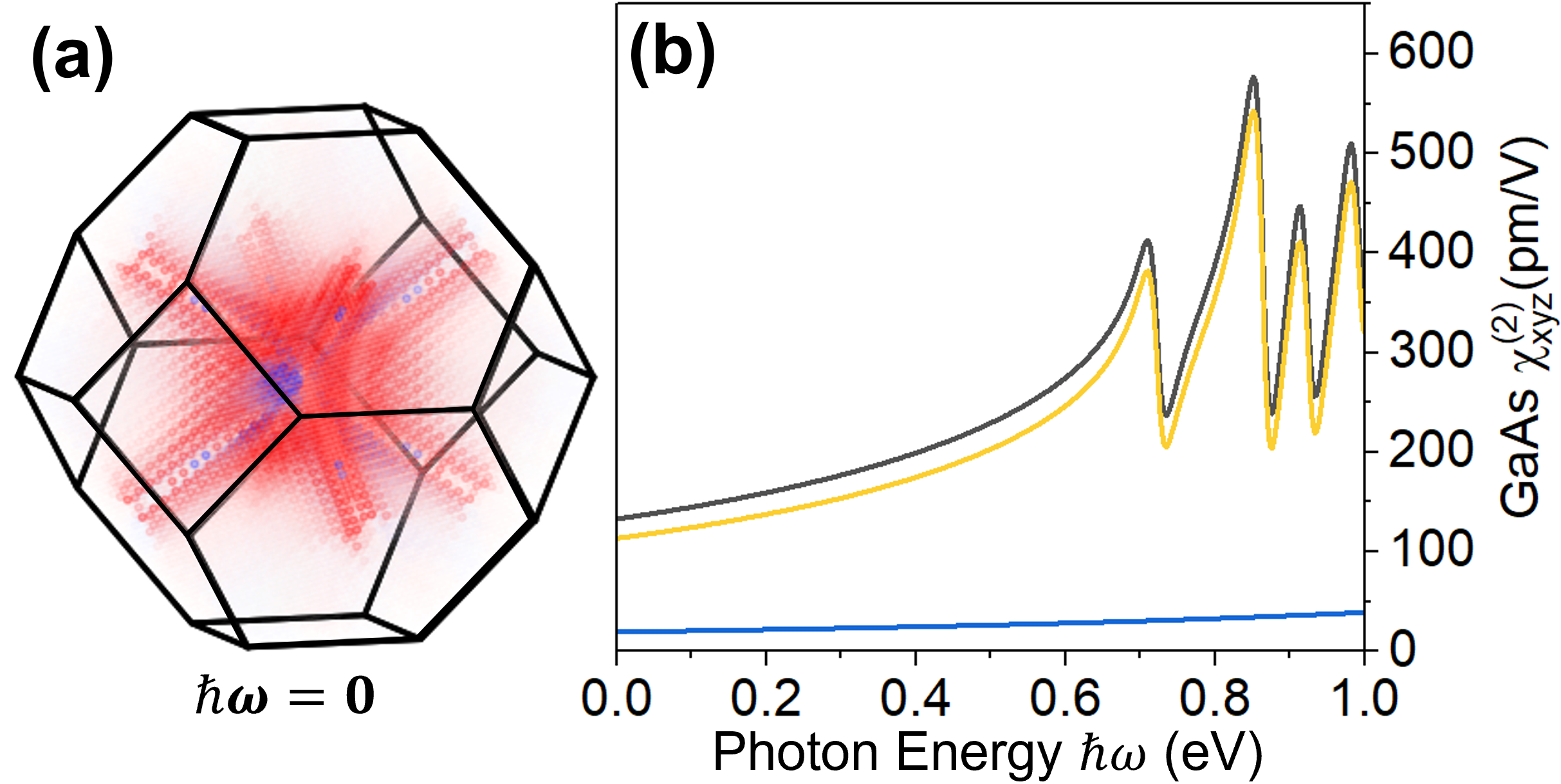}
	\caption{\label{fig:fig6} The calculated nonlinear optical response $\chi^{(2)}$ for GaAs. (a) shows the $k$-dependent contribution to $\chi^{(2)}$ at $\hbar \omega =0$. (b) shows the $\chi^{(2)}(\hbar \omega)$\ where the contribution arising from the entire BZ is shown in black, the contribution arising from the region near $\Gamma$-$L$ (within 0.32 $\AA^{-1}$) is shown in yellow and those from the remainder of the BZ are shown in blue.
	}
\end{figure}				
		
		While such higher order degeneracies have little effect on the linear optical properties, second order harmonics are exclusively determined by the product of matrix elements $\bold{p}_{mn} \bold{p}_{nl} \bold{p}_{lm}$ where m,n, and l are distinct bands. As such, higher order degeneracy becomes paramount as only states which originate from at least a triple degeneracy have substantial contribution to the nonlinear optical response. Taking GaAs as a prototype, it can be seen in Fig S6 that contributions to $\chi^{(2)}$ are dominated by the $\Gamma$-$L$ lines (~85\%), where degenerate FEG planes intersect to form higher order degeneracy. Further, the decomposition of BZ shows that $\chi^{(2)}$ is well represented over the entire energy range by considering only the region of the BZ near the $\Gamma$-$L$ lines.
		

		
		 The key finding of this work is that to a large extent,the wavefunctions of non-transition-metal solids are the same as the FEG, and as such can be represented by a single planewave in most of the BZ. Only in the vicinity of high symmetry areas of the BZ, where the symmetry of the Bravais lattice results in degenerate FEG states, do the wavefunctions of solids have a handful of planewave components. This allows for great simplification of the underlying physics and enables analytical study. Applied to the linear optical response of Si, we find that a single band transition captures the essential physics (nearly 90\% of \(\chi^{(1)}\)). A similar percentage of the nonlinear response for GaAs is found to arise from the higher order degeneracies along the $\Gamma$-$L$ line. The reduced dimensionality, center to this framework, not only greatly improves the understanding of the solid state, empowering inverse design, but also opens the door to drastically reducing the computational effort associated with manybody calculations.



		\section{Acknowledgments}
		This work was supported by the U.S. DOE Grant No. DE-SC0002623. The supercomputer time sponsored by National Energy Research Scientific Center (NERSC) under DOE Contract No. DE-AC02-05CH11231 and the Center for Computational Innovations (CCI) at Rensselaer Polytechnic Institute (RPI) are also acknowledged.  
		
		\bibliographystyle{apsrev4-2}
		\bibliography{Reference}

	\end{document}